\documentclass{ws-ijmpa}
\usepackage[super,compress]{cite}
\usepackage{graphicx}
\usepackage{braket}
\begin{document}
\markboth{Teruyuki Kitabayashi, Hirona Matsumura, Kantaro Minakuchi, Hiroshi Ozaki}{$S_4$ symmetric four-generation models for charged leptons}

\def\Journal#1#2#3#4{{#1} {\bf #2}, #3 (#4)}
\def\ARNPS{Annu. Rev. Nucl. Part. Sci.} 
\def\ANP{Ann. Phys.}
\def\APJ{Astrophys. J.}
\def\APJS{Astrophys. J. Suppl}
\def\COMR{Comptes Rendues}
\def\CPC{Chin. Phys. C}
\def\EPJC{Eur. Phys. J. C}
\def\IJMPA{Int. J. Mod. Phys. A}
\def\IJMPE{Int. J. Mod. Phys. E}
\def\JCAP{JCAP}
\def\JHEP{JHEP}
\def\JETPL{JETP. Lett.}
\def\JETPUSSR{JETP (USSR)}
\def\JPG{J. Phys. G} 
\def\MPLA{Mod. Phys. Lett. A}
\def\NIMA{Nucl. Instrum. Meth. A.}
\def\NATU{Nature}
\def\NCA{Nuovo Cimento}
\def\NJP{New. J. Phys.}
\def\NPB{Nucl. Phys. B}
\def\NPBOLD{Nucl. Phys.}
\def\NPBSUPPL{Nucl. Phys. B. Proc. Suppl.}
\def\PLB{{Phys. Lett.} B}
\def\PMCA{PMC Phys. A}
\def\PREP{Phys. Rep.}
\def\PPNP{Prog. Part. Nucl. Phys.}
\def\PLBOLD{Phys. Lett.}
\def\PAN{Phys. Atom. Nucl.}
\def\PRL{Phys. Rev. Lett.}
\def\PRD{Phys. Rev. D}
\def\PRC{Phys. Rev. C}
\def\PR{Phys. Rev.}
\def\PTP{Prog. Theor. Phys.}
\def\PTEP{Prog. Theor. Exp. Phys.}
\def\RMP{Rev. Mod. Phys.}
\def\SJNP{Sov. J. Nucl. Phys.}
\def\SCIENCE{Science}
\def\TNYAS{Trans. New York Acad. Sci.}
\def\ZETP{Zh. Eksp. Teor. Piz.}
\def\ZFPH{Z. fur Physik}
\def\ZPC{Z. Phys. C}

%
\catchline{}{}{}{}{}
%


\title{$S_4$ symmetric four-generation models for charged leptons}

\author{Teruyuki Kitabayashi\footnote{
teruyuki@tokai-u.jp}
}

\author{Hirona Matsumura\footnote{
8BSNM012@mail.u-tokai.ac.jp}
}

\author{Kantaro Minakuchi\footnote{
8BSNM015@mail.u-tokai.ac.jp}
}

\address{ Department of Physics, Tokai University, 4-1-1 Kitakaname, Hiratsuka, Kanagawa 259-1292, Japan}

\author{Hiroshi Ozaki\footnote{ozaki@tokai-u.jp}}

\address{Laboratory of general education for science and technology, Faculty of Science, Tokai University, 4-1-1 Kitakaname, Hiratsuka, Kanagawa 259-1292, Japan}

\maketitle

\begin{history}
\received{Day Month Year}
\revised{Day Month Year}
\end{history}

\begin{abstract}
We propose $S_4$ symmetric four-generation models for charged leptons. Although an $S_4$ symmetric four-generation model has been already proposed, there are some additional symmetries in the model. We construct four-generation models for charged leptons with only requirement of exact $S_4$ symmetry. It turned out that at least one of the models is consistent with observations of charged lepton masses and predicts the mass of the charged lepton of the fourth generation to be 556 GeV. 
\end{abstract}

\ccode{PACS numbers:11.30.Hv, 12.15.Ff, 12.60.Fr}


\section{Introduction}	

Four-generation models for quarks and leptons are well-motivated extensions of the standard model of the particle physics. These models have been studied extensively in the literature (see Refs \cite{Frampton2000PREP,Holdom2009PMCA} for reviews), for examples, especially for quark sector \cite{Holdom1986PRL,Hatta1994PRD,Chkareuli1999PLB,Hou2010PRD,Smith2012PRD}, for lepton sector \cite{Baer1985PRD,King1992PRD,S-Marcos2002JHEP,Antipin2009JHEP,Frandsen2010PRD,Burdman2010PRD,Deshpande2011PLB}, for interplay between quarks and leptons \cite{Barger1984PRD,Cvetic1995PRD,Djouadi2012PLB,Chen2013PTEP}, for neutrino sector \cite{Gilman1985PRD,Kawasaki1988PRD,Babu1989PLB,Foot1999PRD,Lenz2012PRD,Dutta2016PRD}, for Higgs sector \cite{Belotsky2003PRD,Kribs2007PRD,Ishiwata2011PRD,Bar-Shalom2012PLB,Das2017arXiv} and for dark matter problem  \cite{Raby1988PLB,Belotsky2002PLB,He2010PRD,Lee2011PLB,Zhou2012PRD,Borah2012PRD,Arina2013PLB,Yang2013PRD,Bao2014IJMPA,Hapola2014JCAP,Abdullah2016PRD93,Abdullah2016PRD94}. 

In 1989, Ozaki, one of the authors, proposed an $S_4$ symmetric four-generation model for quarks and charged leptons \cite{Ozaki1989PRD} to extend an $S_4$ symmetric three-generation model \cite{Yamanaka1982PRD}. For $S_4$ models in its early stages, see references in Ref. \cite{Ozaki1989PRD}. Up to now, many three-generation models based on $S_4$ permutation flavor symmetry have been proposed \cite{Brown1984PRD,Ma2006PLB,Hagedron2006JHEP,Altarelli2009JHEP,Grimus2009JPG,Bazzocchi2009PRD,Ishimori2011PRD,Patel2011PLB,Morisi2010PRD,Yang2011PLB,Zhao2011PLB,Vien2015IJMPA,Vien2016IJMPA,Mukherjee2017PRD}; however, four-generation model based on $S_4$ permutation flavor symmetry has not been proposed yet, aside from Ozaki's model.

Although, the predicted values of individual Cabbibo-Kobayashi-Maskawa matrix elements in Ozaki's model were within experimental data in 1989, there are some additional assumptions for the sake of simplicity in calculations and are some additional parameters to fit the model predictions in with observations. These redundant assumptions and additional parameters are disagreeable. Moreover, the charged lepton masses in Ozaki's model were not consistent with currently observed charged lepton masses.

In this paper, we propose exact $S_4$ symmetric four-generation models for charged leptons. It turns out that at least one of the models is consistent with the current observations of the charged lepton masses. 

The paper is organized as follows.  In section \ref{section:setup}, we establish the convention and $S_4$ assignment criteria for model building for later discussions. In section \ref{section:models}, we show the viable model which is consistent with observations. Finally, we give the summary in section \ref{section:summary}.

\section{Setup\label{section:setup}}
\subsection{Tensor products \label{subsection:tensor_products}}
$S_4$ group consists of all permutations among four objects, e.g., $e_1$, $e_2$, $e_3$, and $e_4$. The following tensor products of $S_4$
\begin{eqnarray}
{\bf 3} \otimes {\bf 3} &=& {\bf 3'} \otimes {\bf 3'}={\bf 1} \oplus {\bf 3} \oplus {\bf 2}\oplus {\bf 3'}, \nonumber \\
{\bf 3} \otimes {\bf 3'} &=& {\bf 1'} \oplus {\bf 3'} \oplus {\bf 2}\oplus {\bf 3}, \nonumber \\
{\bf 3} \otimes {\bf 2} &=& {\bf 3'} \otimes {\bf 2} = {\bf 3} \oplus {\bf 3'}, \nonumber \\
{\bf 2} \otimes {\bf 2} &=& {\bf 1} \oplus {\bf 2} \oplus {\bf 1'},
\end{eqnarray}
with obvious products ${\bf 1} \otimes {\bf 1} = {\bf 1^\prime} \otimes {\bf 1^\prime} = {\bf 1}$, ${\bf 1} \otimes {\bf 1^\prime} = {\bf 1'}$, ${\bf 1} \otimes {\bf 3} = {\bf 1'} \otimes {\bf 3'} = {\bf 3}$, ${\bf 1} \otimes {\bf 3'} = {\bf 1'} \otimes {\bf 3} = {\bf 3'}$, ${\bf 1} \otimes {\bf 2} = {\bf 1'} \otimes {\bf 2} = {\bf 2}$ are independent of the basis \cite{Itzykson1966RMP,Ishimori2010PTP}, where ${\bf 1, 1', 2, 3}$ and ${\bf 3'}$ denote the dimensions of irreducible representations of $S_4$ (the prime means antisymmetric representation). We use the basis in Refs.\cite{Yamanaka1982PRD,Ozaki1989PRD}
\begin{eqnarray}
x_1 &=& \frac{1}{2}(e_1+e_2+e_3+e_4), \nonumber \\
x_2 &=& \frac{1}{\sqrt{2}}(e_1-e_2), \nonumber \\
x_3 &=& \frac{1}{\sqrt{2}}(e_3-e_4), \nonumber \\
x_4 &=& \frac{1}{2}(e_1+e_2-e_3-e_4),
\end{eqnarray}
where $x_1$ is a trivial one-dimensional representation ${\bf 1}$ and $\left\{ x_2, x_3, x_4 \right\}$ is a three-dimensional representation ${\bf 3}$. The relevant multiplication rules to our study are as follows:
\begin{eqnarray}
 \left(
  \begin{array}{c}
     a_1  \\
     a_2  \\
  \end{array}
  \right)_{\bf 2}
   \otimes
  \left(
    \begin{array}{c}
     b_1  \\
     b_2  \\
  \end{array}
  \right)_{\bf 2}
 &=&
 \frac{1}{\sqrt{2}}(a_1b_1+a_2b_2)_{\bf 1} \oplus
  \frac{1}{\sqrt{2}}
 \left(
  \begin{array}{c}
     a_1b_2+a_2b_1\\
     a_1b_1-a_2b_2\\
  \end{array}
  \right)_{\bf 2} \nonumber \\
  && \oplus 
   \frac{1}{\sqrt{2}}  (a_1b_2-a_2b_1)_{\bf 1'},
  \end{eqnarray}
\begin{eqnarray}
 \left(
  \begin{array}{c}
     a_1  \\
     a_2  \\
     a_3  \\
  \end{array}
  \right)_{\bf 3}
   \otimes
  \left(
    \begin{array}{c}
     b_1  \\
     b_2  \\
     b_3  \\
  \end{array}
  \right)_{\bf 3}
  &=& \frac{1}{\sqrt{3}}(a_1b_1+a_2b_2+a_3b_3)_{\bf 1}
 \oplus
 \frac{1}{\sqrt{2}}
 \left(
  \begin{array}{c}
     a_1 b_3+a_3b_1\\
     -a_2 b_3-a_3b_2\\
     a_1 b_1-a_2b_2\\
  \end{array}
  \right)_{\bf 3} \nonumber \\
  && \oplus    
   \left(
  \begin{array}{c}
     \frac{1}{\sqrt{2}}(a_1b_2+a_2b_1)  \\
     \frac{1}{\sqrt{6}}(a_1b_1+a_2b_2-2a_3b_3)  \\
  \end{array}
  \right)_{\bf 2} \nonumber \\
  && \oplus  
    \frac{1}{\sqrt{2}} \left(
   \begin{array}{c}
     a_2b_3-a_3b_2  \\
     a_3b_1-a_1b_3\\
     a_1b_2-a_2b_1 \\
  \end{array}
  \right)_{\bf 3'}.
\end{eqnarray}
Several bases of representations of $S_4$ group have been used in the literature \cite{Ishimori2010PTP}.  

\subsection{$S_4$ assignment criteria\label{subsection:S4_assignment_criteria}}
We denote the left-handed (LH) lepton doublets, right-handed (RH) lepton singlets and Higgs doublets by
\begin{eqnarray}
L_i = \left(
  \begin{array}{c}
     \nu_i \\
     \ell_i \\
  \end{array}
  \right)_L, \quad
   E_i=\ell_{iR},
\quad
    \phi_i=\left(
  \begin{array}{c}
     \phi_i^+ \\
     \phi_i^0 \\
  \end{array}
  \right), 
\end{eqnarray}
where $i=1,2,3,4$. The neutrino masses and mixing are omitted in this study \cite{Ozaki1989PRD}.

We assume that these fermions and Higgs scalars are assigned as one of the irreducible representations of $S_4$ according to the following criteria: 
\begin{enumerate}
\item Higgs doublets are assigned as $\phi_1 : {\bf 1}$ and $\left\{ \phi_2, \phi_3, \phi_4 \right\} : {\bf 3}$ in all models. 
\item LH doublets $L_i$ are assigned as ${\bf 1}, {\bf 3}, {\bf 1'}$ or ${\bf 3'}$.
\item RH singlets $E_i$ are assigned as ${\bf 1}, {\bf 2}, {\bf 3}, {\bf 1'}$ or ${\bf 3'}$
\item the order of generations is always $\{1, 2, 3, 4\}$, such as $\{L_1, L_2, L_3\}$, $\{L_2, L_3, L_4\}$, $\{L_2, L_3\}$.
\end{enumerate}

An abbreviation $13$-$112$ will be used to show the following assignment
\begin{eqnarray}
&& L_1 : {\bf 1}, \quad \left\{ L_2,L_3,L_4 \right\}: {\bf 3}, \nonumber \\
&& E_1 : {\bf 1}, \quad  E_2 : {\bf 1}, \quad \left\{ E_3,  E_4 \right\} : {\bf 2},
\end{eqnarray}
with
\begin{eqnarray}
    \phi_1 : {\bf 1}, \quad  \left\{ \phi_2, \phi_3, \phi_4 \right\} : {\bf 3}.
\end{eqnarray}
Other abbreviations $13$-$1'12$, $1'3$-$1'12$, etc., will also be used in the same manner. 

In the next section, we show that at least one of the models (model $1'3$-$11'2$) is consistent with observations.  Before we go to the next section, we show some details of the Higgs potential because the assignment of the Higgs doublets is same in all models.

\subsection{Higgs sector\label{subsection:Higgs_sector}}
The general Higgs potential invariant under $SU(2)\otimes U(1) \otimes S_4$ is \cite{Ozaki1989PRD}
\begin{eqnarray}
V&=&\mu_2^2(\bar{\phi}_2\phi_2 + \bar{\phi}_3\phi_3 +\bar{\phi}_4\phi_4)+\alpha(\bar{\phi}_2\phi_2 + \bar{\phi}_3\phi_3 +\bar{\phi}_4\phi_4)^2 \nonumber \\
&& +\beta\left[\frac{1}{2}(\bar{\phi}_2\phi_3 + \bar{\phi}_3\phi_2)^2 +\frac{1}{6}(\bar{\phi}_2\phi_2 + \bar{\phi}_3\phi_3 -2\bar{\phi}_4\phi_4)^2\right] \nonumber \\
&& +\gamma\left[\frac{1}{2}(\bar{\phi}_2\phi_4 + \bar{\phi}_4\phi_2)^2 +\frac{1}{2}(\bar{\phi}_3\phi_4 + \bar{\phi}_4\phi_3)^2 +\frac{1}{2}(\bar{\phi}_2\phi_2 - \bar{\phi}_3\phi_3)^2\right] \nonumber \\
&& +\delta\left[\frac{1}{2}(\bar{\phi}_2\phi_3 - \bar{\phi}_3\phi_2)^2 +\frac{1}{2}(\bar{\phi}_3\phi_4 - \bar{\phi}_4\phi_3)^2 +\frac{1}{2}(\bar{\phi}_4\phi_2 - \bar{\phi}_2\phi_4)^2\right] \nonumber \\
&& + \mu_1^2\bar{\phi}_1\phi_1 + a \bar{\phi}_1\phi_1 \sum_{i=2}^4 (\bar{\phi}_i\phi_i) + b(\bar{\phi}_1\phi_1)^2 + c\sum_{i=2}^4 \left[(\bar{\phi}_1\phi_i)(\bar{\phi}_1\phi_i) + {\rm h.c.}\right].
\end{eqnarray}

We denote the VEV's of the neutral components of Higgs doublets as 
\begin{eqnarray}
\braket{\phi_1^0}=v_1 e^{-i\vartheta_1}, 
\braket{ \phi_2^0}=v_2 e^{-i\vartheta_2}, \ \braket{ \phi_3^0}=v_3 e^{-i\vartheta_3}, \braket{ \phi_4^0}=v_4 e^{-i\vartheta_4}.
\label{Eq:VEVoriginal}
\end{eqnarray}
\
In terms of the VEV's we have
\begin{eqnarray}
V&=&\mu_2^2(v_2^2 + v_3^2 +v_4^2)+\alpha(v_2^2 + v_3^2 +v_4^2)^2 \nonumber \\
&& +\beta\left[2v_2^2v_3^2\cos^2(\vartheta_2-\vartheta_3)+\frac{1}{6}(v_2^2 + v_3^2-2v_4^2)^2\right] \nonumber \\
&& +\gamma\left[2v_2^2v_4^2\cos^2(\vartheta_2-\vartheta_4) +2v_3^2v_4^2\cos^2(\vartheta_3-\vartheta_4) +\frac{1}{2}(v_2^2 - v_3^2)^2\right] \nonumber \\
&& -\delta\left[2v_3^2v_4^2\sin^2(\vartheta_3-\vartheta_4) +2v_4^2v_2^2\sin^2(\vartheta_4-\vartheta_2) +2v_2^2v_3^2\sin^2(\vartheta_2-\vartheta_3)\right] \nonumber \\
&& + \left\{\mu_1^2 + a  (v_2^2+v_3^2+v_4^2)  + 2 c \left[ v_2^2 \cos 2(\vartheta_2-\vartheta_1) + v_3^2 \cos 2(\vartheta_3-\vartheta_1) \right. \right. \nonumber \\
&& + \left. \left. v_4^2 \cos 2(\vartheta_4-\vartheta_1) \right] \right\} v_1^2 + bv_1^4 .
\end{eqnarray}

The minimization conditions are 
\begin{eqnarray}
\frac{\partial V}{\partial v_1} &=& 2 \left\{ \mu_1^2 +a (v_2^2+v_3^2+v_4^2 )+ 2c\left[v_2^2 \cos 2(\vartheta_2-\vartheta_1) 
+  v_3^2 \cos 2(\vartheta_3-\vartheta_1) \right. \right. \nonumber \\
&& + \left. \left. v_4^2 \cos 2(\vartheta_4-\vartheta_1) \right] \right\}v_1 + 4bv_1^3 =0,
\label{Eq:partial_V/partial_v1}
\end{eqnarray}
\begin{eqnarray}
\frac{\partial V}{\partial v_2} &=&\left(4\alpha + \frac{2}{3}\beta + 2\gamma\right)v_2^3 + \left\{ 2\mu_2^2 + \left[ 4\alpha+4\beta\cos^2(\vartheta_2-\vartheta_3) + \frac{2}{3}\beta \right. \right. \nonumber \\
&& \left. \left.-2\gamma - 4\delta \sin^2(\vartheta_2-\vartheta_3) \right] v_3^2 + \left[ 4\alpha - \frac{4}{3}\beta + 4\gamma \cos^2(\vartheta_2-\vartheta_4) \right. \right. \nonumber \\
&& \left. \left. -4\delta \sin^2(\vartheta_4-\vartheta_2) \right]v_4^2   2av_1^2 + 4cv_1^2 \cos 2(\vartheta_2-\vartheta_1)  \right\} v_2 = 0,
\label{Eq:partial_V/partial_v2}
\end{eqnarray}
\begin{eqnarray}
\frac{\partial V}{\partial v_3} &=&\left(4\alpha + \frac{2}{3}\beta + 2\gamma\right)v_3^3 + \left\{ 2\mu_2^2 +\left[ 4\alpha+4\beta\cos^2(\vartheta_2-\vartheta_3)  + \frac{2}{3}\beta - 2\gamma \right. \right. \nonumber \\
&& - \left. \left. 4\delta \sin^2(\vartheta_2-\vartheta_3) \right]v_2^2 + \left[ 4\alpha - \frac{4}{3}\beta + 4\gamma \cos^2(\vartheta_3-\vartheta_4) \right. \right. \nonumber \\
&& - \left. \left. 4\delta \sin^2(\vartheta_3-\vartheta_4) \right]v_4^2  +  2av_1^2  + 4cv_1^2 \cos 2(\vartheta_3-\vartheta_1) \right\}v_3 = 0,
\label{Eq:partial_V/partial_v3}
\end{eqnarray}
\begin{eqnarray}
\frac{\partial V}{\partial v_4} &=&\left(4\alpha + \frac{8}{3}\beta\right)v_4^3 + \left\{ 2\mu_2^2 +\left[ 4\alpha-\frac{4}{3}\beta +4\gamma\cos^2(\vartheta_2-\vartheta_4) \right. \right. \nonumber \\
&& - \left. \left. 4\delta \sin^2(\vartheta_4-\vartheta_2) \right]v_2^2 + \left[ 4\alpha - \frac{4}{3}\beta + 4\gamma \cos^2(\vartheta_3-\vartheta_4) \right. \right. \nonumber \\
&& - \left. \left. 4\delta \sin^2(\vartheta_3-\vartheta_4) \right]v_3^2  +  2av_1^2  + 4cv_1^2 \cos 2(\vartheta_4-\vartheta_1) \right\}v_4 = 0,
\label{Eq:partial_V/partial_v4}
\end{eqnarray}
\begin{eqnarray}
\frac{\partial V}{\partial \vartheta_1} = 4c\left[ v_2^2 \sin 2(\vartheta_2-\vartheta_1) + v_3^2 \sin 2(\vartheta_3-\vartheta_1) + v_4^2 \sin 2(\vartheta_4-\vartheta_1) \right]v_1^2
=0,
\label{Eq:partial_V/partial_varteta_1}
\end{eqnarray}
\begin{eqnarray}
\frac{\partial V}{\partial \vartheta_2} &=& - 2\beta v_2^2 v_3^2 \sin 2(\vartheta_2-\vartheta_3) - 2\gamma v_2^2 v_4^2 \sin 2(\vartheta_2-\vartheta_4) \nonumber \\
&& + 2\delta \left[ v_2^2 v_4^2 \sin 2(\vartheta_4-\vartheta_2) - v_2^2 v_3^2 \sin 2(\vartheta_2-\vartheta_3) \right] -4c v_2^2 \sin 2(\vartheta_2-\vartheta_1) \nonumber \\
&=& 0,
\label{Eq:partial_V/partial_varteta_2}
\end{eqnarray}
\begin{eqnarray}
\frac{\partial V}{\partial \vartheta_3} &=& - 2\beta v_2^2 v_3^2 \sin 2(\vartheta_3-\vartheta_2) - 2\gamma v_3^2 v_4^2 \sin 2(\vartheta_3-\vartheta_4) \nonumber \\
&& + 2\delta \left[ -v_3^2 v_4^2 \sin 2(\vartheta_3-\vartheta_4) -  v_2^2 v_3^2 \sin 2(\vartheta_3-\vartheta_2)\right]  \nonumber \\
&& -4c v_3^2 \sin 2(\vartheta_3-\vartheta_1) = 0,
\label{Eq:partial_V/partial_varteta_3}
\end{eqnarray}
and
\begin{eqnarray}
\frac{\partial V}{\partial \vartheta_4} &=& 2\gamma \left[ v_2^2 v_4^2 \sin 2(\vartheta_2-\vartheta_4) + v_3^2 v_4^2 \sin 2(\vartheta_3-\vartheta_4) \right]  + 2\delta \left[ v_3^2 v_4^2 \sin 2(\vartheta_3-\vartheta_4) \right. \nonumber \\
&& - \left. v_4^2 v_2^2 \sin 2(\vartheta_4-\vartheta_2)\right] -4c v_4^2 \sin 2(\vartheta_4-\vartheta_1) = 0.
\label{Eq:partial_V/partial_varteta_4}
\end{eqnarray}
The Eq.(\ref{Eq:partial_V/partial_varteta_1}) is not independent of Eqs.(\ref{Eq:partial_V/partial_varteta_2})-(\ref{Eq:partial_V/partial_varteta_4}); this reflects the fact that $V$ depends only on three angles; we can set $\vartheta_1=\vartheta_4$. When we substitute Eqs.(\ref{Eq:partial_V/partial_varteta_2})-(\ref{Eq:partial_V/partial_varteta_4}) into Eqs.(\ref{Eq:partial_V/partial_v1})-(\ref{Eq:partial_V/partial_v4}), we get
\begin{eqnarray}
&&\left(4\alpha - \frac{4}{3}\beta + 2\gamma - 2\delta \right)v_2^2 + \left(4\alpha + \frac{8}{3}\beta - 2\gamma - 2\delta \right)v_3^2 + \left(4\alpha - \frac{4}{3}\beta + 2\gamma \right. \nonumber \\
&& \qquad \qquad \quad \left. - 2\delta \right)v_4^2  + 2\mu_2^2 + 2av_1^2 = 0,
\label{Eq:Vmin_conditions1}
\end{eqnarray}
\begin{eqnarray}
&& \left(4\alpha + \frac{8}{3}\beta - 2\gamma - 2\delta \right)v_2^2 + \left(4\alpha - \frac{4}{3}\beta + 2\gamma - 2\delta \right)v_3^2 + \left(4\alpha - \frac{4}{3}\beta + 2\gamma \right. \nonumber \\
&& \qquad \qquad \quad - \left. 2\delta \right)v_4^2  + 2\mu_2^2 + 2av_1^2 = 0,
\label{Eq:Vmin_conditions2}
\end{eqnarray}
\begin{eqnarray}
&& \left(4\alpha - \frac{4}{3}\beta + 2\gamma - 2\delta \right)v_2^2 + \left(4\alpha - \frac{4}{3}\beta + 2\gamma - 2\delta \right)v_3^2 + \left(4\alpha + \frac{8}{3}\beta \right. \nonumber \\
&& \qquad - \left. 2\frac{(\gamma+\delta)^2}{\beta+\delta} \right)v_4^2 + 2\mu_2^2 + 2av_1^2 - 4c\frac{\gamma+\delta}{\beta+\delta}v_1^2 = 0,
\label{Eq:Vmin_conditions3}
\end{eqnarray}
and
\begin{eqnarray}
2a \left(v_2^2+v_3^2+v_4^2\right) + 2\mu_1^2 - 4c\left(\frac{2c}{\beta+\delta}v_1^2- \frac{\beta-\gamma}{\beta+\delta}v_4^2 \right) + 4 b v_1^2 = 0.
\label{Eq:Vmin_conditions4}
\end{eqnarray}
Eq.(\ref{Eq:Vmin_conditions1}) and Eq.(\ref{Eq:Vmin_conditions2}) imply $v_2^2=v_3^2$. It should be noticed that $v_2^2=v_3^2=v_4^2$ is not necessarily satisfied. The relation of  $v_2^2=v_3^2$ immediately leads to $\sin 2(\vartheta_2-\vartheta_4) + \sin 2(\vartheta_3-\vartheta_4)=0$ or $\vartheta_3-\vartheta_4 = \vartheta_4-\vartheta_2+n\pi$ ($n=0, \pm 1, \pm2,\cdots$) with the help of Eq.(\ref{Eq:partial_V/partial_varteta_4}). We choose the symmetry breaking direction as $\vartheta_1=\vartheta_4=0$ and obtain $\vartheta_3 = -\vartheta_2+n\pi$.
Thus there remains only one phase $\vartheta_2$. Substituting $v_2=v_3=\xi/\sqrt{2}$, $\vartheta_2=\phi$ (and $\vartheta_1 = \vartheta_4 = 0$) into Eq.(\ref{Eq:VEVoriginal}), the VEV's of the neutral components of Higgs doublets are represent as 
\begin{eqnarray}
\braket{\phi_1^0}=v_1, \quad
\braket{ \phi_2^0}=\frac{\xi}{\sqrt{2}}e^{i\phi}, \quad 
\braket{ \phi_3^0}=\frac{\xi}{\sqrt{2}}e^{-i\phi}, \quad 
\braket{ \phi_4^0}=v_4.
\end{eqnarray}
We rewrite Eq.(\ref{Eq:partial_V/partial_varteta_2}) as
\begin{eqnarray}
\sin 4\phi =2\left[ -\frac{\gamma+ \delta}{\beta + \delta} \left( \frac{v_4}{\xi}\right)^2  -\frac{2c}{\beta+\delta}\left( \frac{v_1}{\xi}\right)^2 \right] \sin 2 \phi,
\end{eqnarray}
and divide it by $\sin 2 \phi \neq 0$, then, the minimization condition of the Higgs potential can be obtained as
\begin{eqnarray}
\cos 2\phi = -\frac{\gamma+ \delta}{\beta + \delta} \left( \frac{v_4}{\xi}\right)^2 - \frac{2c}{\beta+\delta}\left( \frac{v_1}{\xi}\right)^2.
\end{eqnarray}
The minimums is stable if
\begin{eqnarray}
b>0, \quad  |\gamma + \delta | > |\beta + \delta |, \quad |2c| > | \beta + \delta |, \quad | v_1 | < |\xi|,  \quad | v_4 | < |\xi|. 
\end{eqnarray}
Moreover, since all four Higgs scalars must contribute to electroweak symmetry breaking, the following sum rule
\begin{eqnarray}
(246~{\rm GeV})^2 = |\braket{\phi_1^0}|^2 + |\braket{ \phi_2^0}|^2 +|\braket{ \phi_3^0}|^2+|\braket{ \phi_4^0}|^2, 
\end{eqnarray}
must be satisfied.

\section{Models\label{section:models}}

\subsection{Candidates\label{subsection:candidates}}
%
\begin{table}[th]
\tbl{Candidates of $S_4$ symmetric four-generation model.}
{\begin{tabular}{@{}cccccccc@{}} \toprule
& model & $L_1\phi_1$ & $L_1\phi_{234}$ & $L_{234}\phi_1$ & $L_{234}\phi_{234}$ & ${\rm rank}(M)$ & \# of couplings \\ \colrule
(a) & $13$-$112$ & $E_1,E_2$ & - & - & $E_1, E_2, E_{34}$ & 4 & 5\\
(b) & $13$-$1'12$ & $E_2$ & - & - & $E_2, E_{34}$ & 3 & 3\\
(c) & $13$-$11'2$ & $E_1$ & - & - & $E_1,E_{34}$ & 3 & 3\\
(d) & $13$-$1'1'2$ & - & - & - & $E_{34}$ & 2 & 1\\
\hline
(e) & $1'3$-$112$ & - & - & - & $E_1, E_2, E_{34}$ & 3 & 3\\
(f) & $1'3$-$1'12$ & $E_1$ & - & - & $E_2, E_{34}$ & 4 & 3\\
(g) & $1'3$-$11'2$ & $E_2$ & - & - & $E_1,E_{34}$ & 4 & 3\\
(h) & $1'3$-$1'1'2$ & $E_1,E_2$ & - & - & $E_{34}$ & 4 & 3\\
\hline
(i) & $13$-$13$ & $E_1$ & $E_{234}$ & $E_{234}$ & $E_1,E_{234}$ & 4 & 5\\
(j) & $13$-$1'3$ & - & $E_{234}$ & $E_{234}$ & $E_{234}$ & 3 & 3\\
(k) & $13$-$13'$ & $E_1$ & - & - & $E_1,E_{234}$ & 4 & 3\\
(l) & $13$-$1'3'$ & - & - & - & $E_{234}$ & 3 & 1\\
\hline
(m) & $1'3$-$13$ & - & - & $E_{234}$ & $E_1,E_{234}$ & 3 & 3\\
(n) & $1'3$-$1'3$ & $E_1$ & - & $E_{234}$ & $E_{234}$ & 4 & 3\\
(o) & $1'3$-$13'$ & - & $E_{234}$ & - & $E_1,E_{234}$ & 4 & 3\\
(p) & $1'3$-$1'3'$ & $E_1$ & $E_{234}$ & - & $E_{234}$ & 4 & 3\\
 \botrule
\end{tabular} \label{Tab:Table1}}
\end{table}

All candidates of $S_4$ symmetric four-generation model for charged leptons satisfied with the $S_4$ assignment criteria in this paper are shown in Table \ref{Tab:Table1}. 

For example, in the line for model (a), $L_1\phi_1E_1$ and $L_1\phi_1E_2$ denote the singlet interactions
\begin{eqnarray}
A_\ell \overline{ (\nu_1,\ell_1)}_L \phi_1 \ell_{1R} + B_\ell \overline{ (\nu_1,\ell_1)}_L \phi_1 \ell_{2R},
\end{eqnarray}
from ${\bf 1} \otimes {\bf 1}\otimes {\bf 1}$, $L_{234}\phi_{234}E_1$ and $L_{234}\phi_{234}E_2$ denote the singlet interactions
\begin{eqnarray}
&& C_\ell \left[ \overline{ (\nu_2,\ell_2)}_L \phi_2 + \overline{ (\nu_3,\ell_3)}_L \phi_3 + \overline{ (\nu_4,\ell_4)}_L \phi_4 \right]\ell_{1R} \nonumber \\
&&+ D_\ell \left[ \overline{ (\nu_2,\ell_2)}_L \phi_2 + \overline{ (\nu_3,\ell_3)}_L \phi_3 + \overline{ (\nu_4,\ell_4)}_L \phi_4 \right]\ell_{2R},
 \end{eqnarray}
from ${\bf 3} \otimes {\bf 3}\otimes {\bf 1}$ and $L_{234}\phi_{234}L_{34}$ denotes the singlet interactions
\begin{eqnarray}
&& E_\ell \left\{ 
\frac{1}{\sqrt{2}}\left[ \overline{ (\nu_2,\ell_2)}_L \phi_3 + \overline{ (\nu_3,\ell_3)}_L \phi_2 \right]\ell_{3R} \right.  \nonumber \\
&& + \left. \frac{1}{\sqrt{6}}\left[ \overline{ (\nu_2,\ell_2)}_L \phi_2 + \overline{ (\nu_3,\ell_3)}_L \phi_3  -  2\overline{ (\nu_4,\ell_4)}_L \phi_4 \right]\ell_{4R}
\right\} ,
\end{eqnarray}
from ${\bf 3} \otimes {\bf 3}\otimes {\bf 2}$. We can see that the rank of mass matrix is equal to four and there are five Yukawa couplings for the charged leptons, $A_\ell , B_\ell , C_\ell , D_\ell , E_\ell $ in the model $13$-$112$. 

There are models in which the symmetric representations are replaced with antisymmetric representation and vice versa (except for Higgs sector). For example, (a) $13$-$112$ yields the following (A) $1'3'$-$1'1'2'$
\begin{eqnarray}
&& L_1 : {\bf 1'}, \quad \left\{ L_2,L_3,L_4 \right\}: {\bf 3'}, \nonumber \\
&& E_1 : {\bf 1'}, \quad  E_2 : {\bf 1'}, \quad \left\{ E_3,  E_4 \right\} : {\bf 2'},
\end{eqnarray}
with 
\begin{eqnarray}
    \phi_1 : {\bf 1}, \quad  \left\{ \phi_2, \phi_3, \phi_4 \right\} : {\bf 3}.
\end{eqnarray}
The structure of the Yukawa Lagrangian in this model $1'3'$-$1'1'2'$ is exactly same as it for the model $13$-$112$. Similarly, (b) $13$-$1'12$ and (B) $1'3'$-$11'2$ have same structure of Yukawa Lagrangian. We omit models (A),(B),(C),$\cdots$,(P) related to (a),(b),(c),$\cdots$,(p) in Table \ref{Tab:Table1}.

\subsection{Viable model ($1'3$-$11'2$) \label{subsection:1'3-11'2}}
We assign the LH doublets, RH singlets and Higgs doublets as ($1'3$-$11'2$) \cite{Ozaki1989PRD}:
\begin{eqnarray}
&& L_1 : {\bf 1'}, \quad \left\{ L_2,L_3,L_4 \right\}: {\bf 3}, \nonumber \\
&& E_1 : {\bf 1}, \quad  E_2 : {\bf 1'}, \quad \left\{ E_3,  E_4 \right\} : {\bf 2},
\end{eqnarray}
and 
\begin{eqnarray}
    \phi_1 : {\bf 1}, \quad  \left\{ \phi_2, \phi_3, \phi_4 \right\} : {\bf 3}.
\end{eqnarray}
The Yukawa Lagrangian relevant for the charged lepton masses and invariant under $SU(2)\otimes U(1) \otimes S_4$ symmetry is
\begin{eqnarray}
\mathcal{L}_Y^\ell&=& A_\ell \overline{ (\nu_1,\ell_1)}_L \phi_1 \ell_{2R} \nonumber \\
&& + B_\ell \left[ \overline{ (\nu_2,\ell_2)}_L \phi_2 + \overline{ (\nu_3,\ell_3)}_L \phi_3 + \overline{ (\nu_4,\ell_4)}_L \phi_4 \right]\ell_{1R}\nonumber \\
&& + C_\ell \left\{ \frac{1}{\sqrt{2}} \left[ \overline{ (\nu_2,\ell_2)}_L \phi_3 + \overline{ (\nu_3,\ell_3)}_L \phi_2 \right]\ell_{3R} \right. \nonumber \\
&& + \left. \frac{1}{\sqrt{6}} \left[ \overline{ (\nu_2,\ell_2)}_L \phi_2 + \overline{ (\nu_3,\ell_3)}_L \phi_3 - 2\overline{ (\nu_4,\ell_4)}_L \phi_4 \right] \ell_{4R} \right\}  \nonumber \\
&& + {\rm h.c.} .
\label{Eq:Yukawa_d_13_112}
\end{eqnarray}

At the tree level, the mass terms in the Lagrangian to be 
\begin{eqnarray}
&& \overline{ (\ell_1,\ell_2,\ell_3,\ell_4)}_L M_\ell \left(
  \begin{array}{c}
     \ell_1 \\
     \ell_2 \\
     \ell_3 \\
     \ell_4 \\
  \end{array}
    \right)_R 
 + {\rm h.c.},
\end{eqnarray}
where the charged lepton mass matrix is 
\begin{eqnarray}
M_\ell  = \left(
  \begin{array}{cccc}
     0 & A_\ell v_1 & 0 & 0   \\
     \frac{B_\ell \xi}{\sqrt{2}} e^{i\phi} & 0 & \frac{C_\ell}{\sqrt{2}} \frac{\xi}{\sqrt{2}} e^{-i \phi} & \frac{C_\ell}{\sqrt{6}} \frac{\xi}{\sqrt{2}} e^{i \phi}   \\
     \frac{B_\ell \xi}{\sqrt{2}} e^{-i \phi} & 0 & \frac{C_\ell}{\sqrt{2}} \frac{\xi}{\sqrt{2}} e^{i \phi} & \frac{C_\ell}{\sqrt{6}} \frac{\xi}{\sqrt{2}} e^{-i \phi} \\
     B_\ell v_4 & 0 & 0 & \frac{-2 C_\ell}{\sqrt{6}} v_4 \\
  \end{array}
  \right), 
\end{eqnarray}

These mass matrices are diagonalized by bi-unitary transformations:
\begin{eqnarray}
U_\ell^\dag M_\ell V_\ell = {\rm diag}.(m_e,m_\mu,m_\tau,m_L),  
\end{eqnarray}
where $m_L$ denotes the masses of charged lepton in fourth generation. The matrices $U_\ell$ and $V_\ell$ for charged lepton sector are satisfied with
\begin{eqnarray}
U_\ell^\dag(M_\ell M_\ell^\dag)U_\ell &=& U_\ell^\dag M_\ell(V_\ell V_\ell^\dag)M_\ell^\dag U_\ell \nonumber \\
&=&(U_\ell^\dag M_\ell V_\ell) (V_\ell^\dag M_\ell^\dag U_\ell) \nonumber \\
&=& (U_\ell^\dag M_\ell V_\ell) (U_\ell^\dag M_\ell V_\ell)^\dag  \nonumber \\
&=& D D^\dag, 
\end{eqnarray}
and
\begin{eqnarray}
V_\ell^\dag(M_\ell^\dag M_\ell)V_\ell &=& V_\ell^\dag M_\ell^\dag(U_\ell U_\ell^\dag)M_\ell V_\ell \nonumber \\
 &=& (V_\ell^\dag M_\ell^\dag U_\ell) (U_\ell^\dag M_\ell V_\ell) \nonumber \\
 &=& (U_\ell^\dag M_\ell V_\ell)^\dag (U_\ell^\dag M_\ell V_\ell) \nonumber \\
 &=& D^\dag D, 
\end{eqnarray}
where $D$ denotes a diagonal mass matrix. 

We have performed a  parameter search and found that, in the model $1'3$-$11'2$, the parameter set
\begin{eqnarray}
v_1 = 73.8~{\rm GeV}, \quad \xi = 229.7~{\rm GeV}, \quad \phi=1.14 \times 10^{-3} ~{\rm rad}, 
\end{eqnarray}
and
\begin{eqnarray}
A_\ell=2.41 \times 10^{-2} , \quad B_\ell= 2.37, \quad C_\ell=1.59 \times 10^{-3}, 
\end{eqnarray}
yields the following masses of charged leptons
\begin{eqnarray}
&& m_e=0.511~{\rm MeV}, \quad m_\mu=106~{\rm MeV}, \nonumber \\
&& m_\tau = 1.78~{\rm GeV}, \quad m_L=556~{\rm GeV}.
\end{eqnarray}
These predicted masses are consistent with the charged lepton masses in these predicted masses PDG\cite{PDG}:
\begin{eqnarray}
&& m_e^{\rm PDG}=0.511~{\rm MeV}, \quad m_\mu^{\rm PDG}=106~{\rm MeV} , \nonumber \\
&& m_\tau^{\rm PDG} = 1.778~{\rm GeV}, \quad m_L^{\rm PDG} \gtrsim  100~{\rm GeV}.
\end{eqnarray}
The $1'3$-$11'2$ model is a viable four-generation model for charged leptons based on exact $S_4$ symmetry.

We would like to note that we have constructed {\it exact} $S_4$ symmetric four-generation model for charged leptons. There are excellent $S_4$ based three-generation models with some additional symmetries such as in Ref.\cite{Morisi2010PRD,Vien2016IJMPA,Mukherjee2017PRD}. These models are valuable because not only charged lepton sector but also quark sector are correctly described in these models. On the other hand, we construct four-generation models only for charged leptons but with only requirement of exact $S_4$ symmetry (we want to know how much we can construct four-generation models for charged leptons based on {\it only} $S_4$ symmetry). Our approach could be one of the way to study $S_4$ symmetric models.

The numerical calculations for other models have been performed. From our numerical calculations, it seems that the model $1'3$-$11'2$ is only viable four-generation models for charged leptons based on exact $S_4$ permutation symmetry.

\section{Summary\label{section:summary}}
We have proposed $S_4$ symmetric four-generation models for charged leptons. Although an $S_4$ symmetric four-generation model has been already proposed\cite{Ozaki1989PRD}, there are some redundant assumptions and additional parameters in the model. In this paper, we have constructed models with only requirement of exact $S_4$ symmetry. We have shown that at least one of the models (model $1'3$-$11'2$) is consistent with observations for masses of charged leptons and predicts the mass of the charged lepton of the fourth generation to be 556 GeV. 

We comment that there is the flavor changing neutral current (FCNC) problem\cite{Ozaki1989PRD} in the model $1'3$-$11'2$. We expect that the FCNC problem may be solved under an assumption that the neutral members are superheavy\cite{Casalbuoni1988NPB}. 

Finally, we note that there are interesting but unsolved matters in the models in this paper, such as (1) including neutrino masses and mixings, (2) considerations for neutral fourth generation particle as a dark matter, (3) effects on collider phenomenology, etc. More details of these topics will be found in our future study.

\vspace{3mm}







\end{document}